\documentclass[prd,showpacs,twocolumn,
amsmath,amssymb,superscriptaddress,floatfix,nofootinbib]{revtex4-1}
\usepackage[utf8]{inputenc}
\usepackage[sort&compress]{natbib}
\usepackage[normalem]{ulem}
\usepackage{lipsum} 
\usepackage{bm}
\usepackage{times}
\usepackage{amssymb,amsbsy,amsmath,amsfonts}
\usepackage{graphicx}
\usepackage{float}
\usepackage{color}
\usepackage{morefloats}
\usepackage{rotating}
\usepackage{srcltx}
\usepackage{slashed}
\usepackage{subfigure}
\usepackage{multirow}
\usepackage{verbatim}
\usepackage{hyperref}
\usepackage{tabularx}
\usepackage{adjustbox}
\usepackage[dvipsnames]{xcolor}
\usepackage{nicefrac}

\usepackage{hyperref}
\hypersetup{
	colorlinks	=true,
	urlcolor	=blue,
	linkcolor	=blue,
	citecolor	=blue,
	pdftitle	={},
	pdfauthor	={Zhi-Wei Liu, Jun-Xu Lu, Ming-Zhu Liu, Li-Sheng Geng},
	pdfsubject	={$\Sigma_c\bar{D}^{(*)}$ correlation function}
	}

\begin{document}

\title{
Distinguishing the spins of $P_c(4440)$ and $P_c(4457)$ with femtoscopic correlation functions
}

\author{Zhi-Wei Liu}
\affiliation{School of Physics, Beihang University, Beijing 102206, China}

\author{Jun-Xu Lu}
\affiliation{School of Physics, Beihang University, Beijing 102206, China}

\author{Ming-Zhu Liu}
\affiliation{School of Space and Environment, Beihang University, Beijing 102206, China}
\affiliation{School of Physics, Beihang University, Beijing 102206, China}

\author{Li-Sheng Geng}
\email[Corresponding author: ]{lisheng.geng@buaa.edu.cn}
\affiliation{School of Physics, Beihang University, Beijing 102206, China}
\affiliation{Peng Huanwu Collaborative Center for Research and Education, Beihang University, Beijing 100191, China}
\affiliation{Beijing Key Laboratory of Advanced Nuclear Materials and Physics, Beihang University, Beijing 102206, China }
\affiliation{Southern Center for Nuclear-Science Theory (SCNT), Institute of Modern Physics, Chinese Academy of Sciences, Huizhou 516000, Guangdong Province, China}

\begin{abstract}
The spins of the pentaquark states $P_c(4440)$ and $P_c(4457)$ play a decisive role in unraveling their nature, but remain undetermined experimentally. Assuming that they are $\Sigma_c\bar{D}^{*}$ bound states, we demonstrate how one can determine their spins by measuring the $\Sigma_c^+\bar{D}^{(*)0}$ correlation functions. We show that one can use the $\Sigma_c^+\bar{D}^0$ correlation function to fix the size of the Gaussian source and then determine the strength of the $\Sigma_c^+\bar{D}^{*0}$ interaction of spin $1/2$ and $3/2$ and therefore the spins of the $P_c(4440)$ and $P_c(4457)$ states. The method proposed can be applied to decipher the nature of other hadronic molecules and thus deepen our understanding of the nonperturbative strong interaction. 

\end{abstract}


\maketitle

\section{Introduction}
In 2015, the LHCb Collaboration observed two pentaquark states, $P_c(4380)$ and $P_c(4450)$, in the $J/\psi p$ invariant mass distribution of the $\Lambda_{b}\to J/\psi p K$ decay~\cite{LHCb:2015yax}. The analysis was updated in 2019 with ten times more data, where the $P_c(4450)$ state was found to split into two states, $P_c(4440)$ and $P_c(4457)$ and in addition a new state $P_c(4312)$ was observed~\cite{LHCb:2019kea}. This is followed by the observation of more pentaquark states containing strangeness~\cite{LHCb:2020jpq} and in $B$ decays~\cite{LHCb:2021chn,LHCb:2022jad}. Their nature as hadronic molecules of $\Sigma_c\bar{D}^{(*)}$~\cite{Chen:2019asm,He:2019ify,Liu:2019tjn,Chen:2019bip,Xiao:2019mvs,Xiao:2019aya,Meng:2019ilv,Yamaguchi:2019seo,Burns:2019iih,PavonValderrama:2019nbk,Du:2019pij,Wang:2019spc,Liu:2019zvb,Lin:2019qiv,Xu:2020gjl,Peng:2020gwk,Yalikun:2021bfm,Xie:2022hhv}, compact multiquark states $c\bar{c}qqq$~\cite{Ali:2019npk,Mutuk:2019snd,Wang:2019got,Cheng:2019obk,Weng:2019ynv,Zhu:2019iwm,Pimikov:2019dyr,Ruangyoo:2021aoi,Deng:2022vkv}, hadron-charmonia~\cite{Eides:2019tgv} or even threshold effects and triangle singularities~\cite{Nakamura:2021qvy} have been proposed and studied in detail, but no firm conclusion has been reached. 

An interesting and key issue related to the two pentaquark states $P_c(4440)$ and $P_c(4457)$ is their spins, which can help distinguish various theoretical interpretations. In the $\Sigma_{c}\bar{D}^*$ molecular picture, their spin can be either $1/2$ or $3/2$, which leads to two scenarios referred to as A and B in Ref.~\cite{Liu:2019tjn}.
In the one boson exchange (OBE) model, if one includes the delta term of the spin-spin potential, scenario A is preferred~\cite{Chen:2019asm}, while if the delta potential is neglected, scenario B is preferred~\cite{Liu:2019zvb}. Recently Yalikun \textit{et al}. characterized the delta potential with a parameter and their results favored scenario B~\cite{Yalikun:2021bfm}. Later, Zhang \textit{et al}. found that a neural network-based approach can discriminate their spins~\cite{Zhang:2023czx}. In Ref.~\cite{Pan:2019skd}, it is argued that one can study the dibaryon states of $\Sigma_c\Xi_{cc}$ to determine the spins of $P_c(4440)$ and $P_c(4457)$ because the dibaryon system and the pentaquark system are related with each other via the heavy antiquark diquark symmetry. 
We note that there have been a large number of theoretical studies trying to distinguish their spins from various perspectives, e.g., their masses~\cite{Liu:2019tjn,Meng:2019ilv,PavonValderrama:2019nbk}, the invariant mass distributions~\cite{Du:2019pij,Du:2021fmf,Burns:2022uiv}, their two-body~\cite{Xiao:2019mvs,Lin:2019qiv}, three-body~\cite{Xie:2022hhv} and radiative~\cite{Ling:2021lmq} decays, magnetic momenta~\cite{Ozdem:2021ugy,Li:2021ryu,Guo:2023fih}, and even their production rates in the $\Lambda_b$ decay~\cite{Wu:2019rog} or in the inclusive processes~\cite{Wang:2019krd,Xie:2020niw,Ling:2021sld,Shi:2022ipx}. Clearly, a determination of the spins of the pentaquark states is extremely important to verify their nature. 

In recent years, femtoscopy, which measures two-hadron momentum correlation functions (CFs) in high-energy collisions, has made remarkable progress in probing the strong interactions between various hadron pairs~\cite{STAR2015Nature527.345, STAR2015PRL114.022301, HADES2016PRC94.025201, ALICE2017PLB774.64, ALICE2017PRC96.064613, ALICE2019PRL123.112002, ALICE2019PLB797.134822, ALICE2019PLB790.22, STAR2019PLB790.490, ALICE2019PRC99.024001, ALICE2020Nature588.232, ALICE2020PRL124.092301, ALICE2020PLB802.135223, ALICE2020PLB805.135419, ALICE2021PRL127.172301, ALICE2021PLB813.136030, ALICE2021PRC103.055201, ALICE2022PLB829.137060, ALICE2022PLB833.137272, ALICE2022PLB833.137335, ALICE2022PRD106.052010}. In particular, it was shown that CFs can help reveal the existence of bound states~\cite{Ohnishi2017PPNP95.279, ALICE2021ARNPS71.377}. In this sense, femtoscopy offers a valuable means to directly test the hadronic molecular picture. It is worthwhile to note~\footnote{In this work, charge conjugated states are always implied unless otherwise stated.} that the $DD^*$ and $D\bar{D}^*$ CFs have been employed to test the nature of $T_{cc}^+(3875)$ and $X(3872)$ as bound states of $DD^*$ and $D\bar{D}^*$~\cite{Kamiya:2022thy, Vidana:2023olz}. In three recent works, the CFs of the open-charm $D_{(s)}\phi$ (where $\phi$ refers to a Nambu-Goldstone boson) pairs in the $(S,I)=(1,0)$ and $(0,1/2)$ sectors and the related $D_{s0}^*(2317)$ as well as $D_0^*(2300)$ states were studied~\cite{Liu:2023uly, Albaladejo:2023pzq, Ikeno:2023ojl}. Based on the square-well model, Ref.~\cite{Liu:2023uly} revealed some general features of CFs. For a moderately attractive potential capable of generating a shallow bound state, the low-momentum CF is above unity for a small size source while below unity for a large size one. On the other hand, for a strongly attractive potential that can generate a deep bound state, it remains between zero and unity for sources of all possible sizes. In addition, the recent measurement of the $pD^-$ CF by the ALICE collaboration demonstrated that it is possible to access the charm sector in experiments~\cite{ALICE:2022enj}.

The measured CFs are usually spin-averaged and therefore can not be employed to directly probe the spin configurations of hadron-hadron interactions. Nevertheless together with other constraints, one can deduce the spin dependence of the underlying hadron-hadron interactions from the spin-averaged CFs~\cite{Haidenbauer:2021zvr, Chizzali:2022pjd}. For instance, a measurement of the $\Sigma^+p$ CF combined with the scattering cross-section allows one to pin down the spin singlet and triplet $S$-wave $\Sigma N$ interactions~\cite{Haidenbauer:2021zvr}. After fixing the spin $3/2$ channel with the lattice QCD simulations~\cite{Lyu:2022imf}, the spin $1/2$ channel of the $p\phi$ (where $\phi$ refers to a ground-state vector meson) interaction, which is strong enough to support the formation of a bound state, was determined by a constrained fit to the experimental $p\phi$ CF~\cite{Chizzali:2022pjd}. In the present work, assuming that $P_{c}(4440)$ and $P_c(4457)$ are deep and shallow bound states of $\Sigma_c\bar{D}^*$, respectively, we propose to determine the spins of $P_c(4440)$ and $P_c(4457)$ by measuring the $\Sigma_c^+\bar{D}^{(*)0}$ CFs.

The manuscript is organized as follows: In Sec. II we provide a brief overview of the formalism to determine the $\Sigma_c\bar{D}^{(*)}$ interactions and calculate their CFs. In Sec. III, we explain how to use the $\Sigma_c^+\bar{D}^{(*)0}$ CFs to deduce the $\Sigma_c\bar{D}^*$ interactions of spin 1/2 and 3/2 and therefore help distinguish the spins of $P_c(4440)$ and $P_c(4457)$. We end with a short summary.

\section{Theoretical Framework}\label{Sec:framework}

In this section, we briefly recall the formalism to derive the $\Sigma_c\bar{D}^{(*)}$ interactions and to calculate the corresponding femtoscopic CFs. In the present work, to derive the $\Sigma_c\bar{D}^{(*)}$ interactions, we turn to the resonance saturation method~\cite{Ecker:1988te}. We note that very recently the spectra of hadronic molecules composed of charmed and anti-charmed hadrons have been systemically investigated in this approach, yielding results consistent with those of the majority of other approaches~\cite{Dong:2021bvy, Dong:2021juy,Peng:2021hkr,Peng:2020xrf}.

Following Ref.~\cite{Dong:2021juy}, the $I=3/2$ and $I=1/2$ $\Sigma_c\bar{D}^{(*)}$ interactions can be described by two contact terms at leading order, assuming that they are saturated by the vector meson exchanges, i.e.,
\begin{subequations}\label{Eq:V_isospin}
\begin{align}
  &\widetilde{V}_{11}^{I=\frac{3}{2}}=2M_{\Sigma_c}M_{\bar{D}^{(*)}}\widetilde{\beta}_1\widetilde{\beta}_2g_V^2\left(\frac{1}{m_\omega^2}+\frac{1}{m_\rho^2}\right),\\
  &\widetilde{V}_{22}^{I=\frac{1}{2}}=2M_{\Sigma_c}M_{\bar{D}^{(*)}}\widetilde{\beta}_1\widetilde{\beta}_2g_V^2\left(\frac{1}{m_\omega^2}-\frac{2}{m_\rho^2}\right),
\end{align}
\end{subequations}
where $M_{\Sigma_c}$ ($M_{\bar{D}^{(*)}}$) and $m_\rho$ ($m_\omega$) are the isospin-averaged masses of the heavy hadrons and the exchanged particles, respectively. Assuming vector meson dominance, the coupling constants $g_V$, $\widetilde{\beta}_1$ and $\widetilde{\beta}_2$ are estimated to be $5.8$~\cite{Bando:1987br}, $0.9$~\cite{Isola:2003fh} and $1.74/2$~\cite{Liu:2011xc, Chen:2019asm}, accordingly. To compare with future femtoscopy experiments, we have to work in charge basis. 
The potentials in charge basis can be expressed in terms of those in isospin basis as follows,
\begin{subequations}\label{Eq:V_particle}
\begin{align}
  V_{11}&=\frac{2}{3}\widetilde{V}_{11}^{I=\frac{3}{2}}+\frac{1}{3}\widetilde{V}_{22}^{I=\frac{1}{2}},\\
  V_{22}&=\frac{1}{3}\widetilde{V}_{11}^{I=\frac{3}{2}}+\frac{2}{3}\widetilde{V}_{22}^{I=\frac{1}{2}},\\
  V_{12}&=V_{21}=\frac{\sqrt{2}}{3}\widetilde{V}_{11}^{I=\frac{3}{2}}-\frac{\sqrt{2}}{3}\widetilde{V}_{22}^{I=\frac{1}{2}},
\end{align}
\end{subequations}
where the indices $\nu(\nu')=1, 2$ represent the $\Sigma_c^{+}\bar{D}^{(*)0}$ and $\Sigma_c^{++}D^{(*)-}$ channels, respectively. Note that both the $\widetilde{V}_{\mu'\mu}$ and $V_{\nu'\nu}$ potentials above are in $S$-wave, and do not contain any free parameter.

The effects of final-state interactions are encoded in the relative wave function of the hadron pair of interests, which is one of two essential components of the femtoscopic CF~\cite{Ohnishi2017PPNP95.279, ALICE2021ARNPS71.377}. In general, there are two ways to obtain the scattering wave function, either solving the Schr\"odinger equation in coordinate space~\cite{ALICE2018EPJC78.394, Ohnishi2021PRC105.014915}, or the Lippmann-Schwinger (Bethe-Salpeter) equation in momentum space~\cite{Haidenbauer2019NPA981.1, Liu2022CPC}. In the present work the $\Sigma_c\bar{D}^{(*)}$ potentials are constructed in momentum space, therefore it is more convenient to first obtain the reaction amplitude $T$ by solving the scattering equation $T=V+VGT$, and then derive the scattering wave function in the center-of-mass frame using the relation $|\psi\rangle=|\varphi\rangle+GT|\varphi\rangle$, where $G$ and $|\varphi\rangle$ represent the free propagator and the free wave function, respectively. More specifically, similar to our recent works~\cite{Liu2022CPC, Liu:2023uly}, we use the following coupled-channel scattering equation to obtain the reaction amplitude,
\begin{align}\label{Eq:Kadyshevsky}
  &T_{\nu'\nu}(k',k)=V_{\nu'\nu}\cdot f_{\Lambda_F}(k',k)+\sum_{\nu^{\prime\prime}}\int_0^\infty\frac{{\rm d}k^{\prime\prime}k^{\prime\prime2}}{8\pi^2}\nonumber\\
  &\times\frac{V_{\nu'\nu^{\prime\prime}}\cdot f_{\Lambda_F}(k',k^{\prime\prime})\cdot T_{\nu^{\prime\prime}\nu}(k^{\prime\prime},k)}{E_{\Sigma_c,\nu^{\prime\prime}}E_{\bar{D}^{(*)},\nu^{\prime\prime}}(\sqrt{s}-E_{\Sigma_c,\nu^{\prime\prime}}-E_{\bar{D}^{(*)},\nu^{\prime\prime}}+i\epsilon)},
\end{align}
where $\sqrt{s}=E_{\Sigma_c,\nu}(k)+E_{\bar{D}^{(*)},\nu}(k)$, and $E_{\Sigma_c(\bar{D}^{(*)}),\nu}(k)=\sqrt{k^2+M_{\Sigma_c(\bar{D}^{(*)}),\nu}^2}$. As shown in Eq.~\eqref{Eq:Kadyshevsky}, in order to avoid ultraviolet divergence in numerical evaluations, we multiply the potential $V_{\nu'\nu^{(\prime\prime)}}$ by a Gaussian regulator $f_{\Lambda_F}(k,k')=\exp[-(k/\Lambda_F)^{2}-(k'/\Lambda_F)^{2}]$ to suppress high-momentum contributions~\cite{Liu:2019tjn, Liu:2023uly}, where $\Lambda_F$ is a cutoff parameter to be determined. We can then compute the scattering wave function with the half-off-shell $T$-matrix in the following way,
\begin{align}\label{Eq:Fourier_Bessel}
  &\widetilde{\psi}_{\nu'\nu}(k,r)=\delta_{\nu'\nu}j_0(kr)+\int_0^\infty\frac{{\rm d}k'k^{\prime2}}{8\pi^2}\nonumber\\
  &\times\frac{T_{\nu'\nu}(k',k)\cdot j_0(k'r)}{E_{\Sigma_c,\nu^{\prime}}E_{\bar{D}^{(*)},\nu^{\prime}}(\sqrt{s}-E_{\Sigma_c,\nu^{\prime}}-E_{\bar{D}^{(*)},\nu^{\prime}}+i\epsilon)},
\end{align}
where $j_0$ is the spherical Bessel function of angular momentum $l=0$. 

The other essential component of the CF is the so-called particle-emitting source, which characterizes the distribution of the relative distance $r$ at which the hadron pair of interests is emitted~\cite{Ohnishi2017PPNP95.279, ALICE2021ARNPS71.377}. In this work, we adopt the widely used static and spherical Gaussian source with a single parameter $R$, namely, $S_{12}(r)=\exp[-r^2/(4R^2)]/(2\sqrt{\pi}R)^3$. With the aforementioned two theoretical ingredients, the CF can be calculated with the Koonin-Pratt formula~\cite{Koonin1977PLB70.43, Pratt1990PRC42.2646, Bauer1992ARNPS42.77}
\begin{align}\label{Eq:CF}
  C(k)=1+&\int_0^\infty4\pi r^2{\rm d}rS_{12}(r)\nonumber\\
  &\times\left[\sum_{\nu'}\omega_{\nu'}\left|\widetilde{\psi}_{\nu'\nu}(k,r)\right|^2-\left|j_0(kr)\right|^2\right],
\end{align}
where $\omega_{\nu'}$ is the weight for each individual component of the multichannel wave function, and the sum runs over all possible coupled channels. For simplicity we assume that the weights are the same and equal to $1$ in this exploratory study.

It is worthwhile to note that in Refs.~\cite{Vidana:2023olz, Albaladejo:2023pzq, Ikeno:2023ojl} the authors developed a formalism to factorize the scattering amplitudes outside the integrals and studied the femtoscopic CFs of the $D\bar{D}^*$ and $D_{(s)}\phi$ scattering, where $\phi$ refers to a Nambu-Goldstone boson.

\section{Results and Discussions}
\subsection{Fixing the source size with the $\Sigma_c\bar{D}$ correction function}
We first concentrate on the $\Sigma_c^+\bar{D}^0$ interaction because the $P_c(4312)$ state can be viewed as a bound state of $\Sigma_c\bar{D}$. As explained above, the $\Sigma_c\bar{D}$ interaction obtained in the resonance saturation model does not contain any free parameter, leaving the cutoff momentum as the only unknown parameter to be determined. We thus fine-tune the cutoff $\Lambda_F$ to be $960$ MeV to reproduce the experimental mass of $P_c(4312)$~\cite{LHCb:2019kea}. The masses of the pentaquark states, the thresholds of the $\Sigma_c^+\bar{D}^{(*)0}$ pairs, and the corresponding binding energies are given in Table~\ref{Tab:Pentaquark}.

\begin{table}[htbp]
\centering
\setlength{\tabcolsep}{11.2pt}
\caption{Masses of the three pentaquark states, $\Sigma_c^+\bar{D}^{(*)0}$ thresholds, and the corresponding binding energies $B$ (in units of MeV).}\label{Tab:Pentaquark}
\begin{tabular}{cccc}
  \hline
  \hline
  Pentaquark & Mass & Threshold & $B$  \\ \hline
  $P_c(4312)$ & $4311.9$~\cite{LHCb:2019kea}  & $\Sigma_c^+\bar{D}^{0}~~(4317.5)$  & $~~5.6$  \\ \hline
  $P_c(4440)$ & $4440.3$~\cite{LHCb:2019kea}  & $\Sigma_c^+\bar{D}^{*0}(4459.5)$ & $19.2$  \\ \hline
  $P_c(4457)$ & $4457.3$~\cite{LHCb:2019kea}  & $\Sigma_c^+\bar{D}^{*0}(4459.5)$ & $~~2.2$ \\
  \hline
  \hline
\end{tabular}
\end{table} 

\begin{figure}[htbp]
  \centering
  \includegraphics[width=0.48\textwidth]{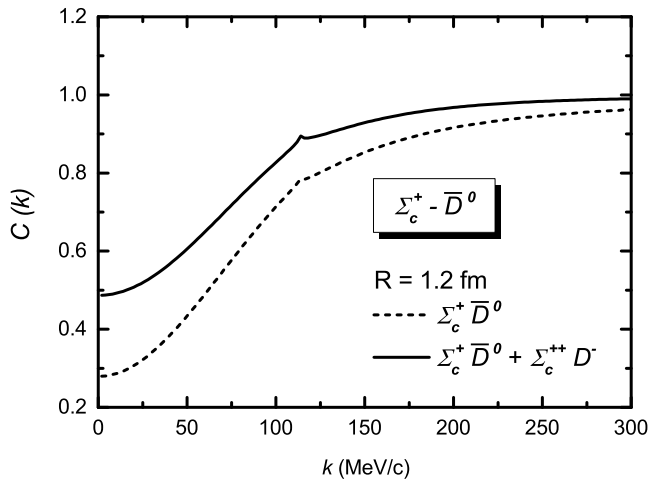}
  \caption{$\Sigma_c^+\bar{D}^0$ correlation function as a function of the relative momentum $k$. The results are obtained with the contact potential and a Gaussian source of $R=1.2$ fm. The dashed line denotes the correlation function for which only the $\Sigma_c^+\bar{D}^0$ contribution is taken into account, while the solid line includes also the $\Sigma_c^+\bar{D}^0-\Sigma_c^{++}D^-$ contribution.
  }\label{Fig:Pc4312_CC}
\end{figure}

The predicted $\Sigma_c^+\bar{D}^0$ CF is shown in Fig.~\ref{Fig:Pc4312_CC}. The results are obtained with the contact potentials derived above and a Gaussian source of $R=1.2$ fm. Obviously, there is a significant suppression in the $\Sigma_c^+\bar{D}^0$ correlation compared to unity in a wide range of the relative momentum $k$, which is consistent with the
low energy behavior of a CF corresponding to a hadron-hadron interaction capable of generating a bound state~\cite{Liu:2023uly}. The contribution from the $\Sigma_c^+\bar{D}^0-\Sigma_c^{++}D^-$ transition leads to a cusplike structure around $k\simeq114$ MeV/c, which corresponds to the opening of the $\Sigma_c^{++}D^-$ channel. In fact, the sizable coupled-channel effect can be traced back to the strong $\Sigma_c^+\bar{D}^0-\Sigma_c^{++}D^-$ off-diagonal interaction. The resonance saturation model~\cite{Dong:2021bvy} adopted in the present work predicts an attractive $\Sigma_c\bar{D}$ interaction for the $I=1/2$ sector but a strongly repulsive $\Sigma_c\bar{D}$ interaction for the $I=3/2$ sector. As a consequence, the off-diagonal interaction in charge basis is even stronger since it is proportional to $\widetilde{V}_{11}^{I=\frac{3}{2}}- \widetilde{V}_{22}^{I=\frac{1}{2}}$. One should keep in mind that we know very little about the $I=3/2$ interaction so far, hence the $\Sigma_c^+\bar{D}^0-\Sigma_c^{++}D^-$ coupling strength may suffer from relatively large uncertainties. Future high precision measurements of the $\Sigma_c^+\bar{D}^0$ CFs, especially near the coupled-channel threshold, would help to determine the $\Sigma_c^+\bar{D}^0-\Sigma_c^{++}D^-$ coupling strength, similar to the studies of the $\Lambda N-\Sigma N$ and $N\bar{K}-\Sigma\pi$ transition strengths by the ALICE Collaboration~\cite{ALICE2022PLB833.137272, ALICE:2022yyh}.

\begin{figure}[htbp]
  \centering
  \includegraphics[width=0.48\textwidth]{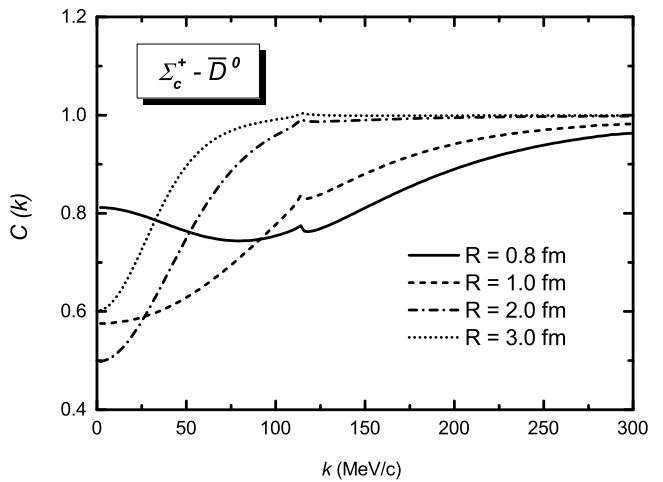}
  \caption{Source size ($R$) dependence of the $\Sigma_c^+\bar{D}^0$ correlation function. 
  }\label{Fig:Pc4312_SS}
\end{figure}

In Fig.~\ref{Fig:Pc4312_SS}, we show the source size dependence of the $\Sigma_c^+\bar{D}^0$ CF. For both small and large emitting sources, the $\Sigma_c^+\bar{D}^0$ CFs are all between zero and unity, which is similar to the case of the $D^0K^+$ system~\cite{Liu:2023uly}. However, because the binding energy of the $P_c(4312)$ state is only $5.6$ MeV, the $\Sigma_c^+\bar{D}^0$ CFs in the low-momentum region do not decrease monotonically with the decreasing source size. The size dependence is reversed in the small collision system, which is close to the case of shallow bound states. The obvious and nonmonotonic source size dependence indicates that the $\Sigma_c^+\bar{D}^0$ CF can be utilized to determine the source size.

Since the spin of the $\Sigma_c$ is $1/2$ and that of the $\bar{D}$ is $0$, only the spin $1/2$ configuration can be probed in the $\Sigma_c^+\bar{D}^0$ CF. Once the $\Sigma_c\bar{D}$ interaction is determined reasonably as done in this work, the space-time dimension of the emitting source is the only unknown factor in the femtoscopic study, which means that the size of the Gaussian source can be inferred from the measurement of the $\Sigma_c^+\bar{D}^0$ CF. In other words, the femtoscopic study of the $\Sigma_c^+\bar{D}^0$ CF can help fix the source size, which is similar to the measurement of the proton-proton CF~\footnote{The nuclear force is known with high precision.} in studying the hyperon-nucleon and hyperon-hyperon interactions~\cite{ALICE2019PRL123.112002, ALICE2019PLB797.134822}. Previous studies on the sizes of sources creating various particle pairs found that by accounting for the effects of strong decays, a common (core) source can be expressed as a function of the transverse mass~\cite{ALICE:2020ibs}. In fact, this core-resonance model has been successfully employed in several femtoscopic analyses, such as $p\Omega^-$~\cite{ALICE2020Nature588.232}, $p\phi$~\cite{ALICE2021PRL127.172301} and $pD^-$~\cite{ALICE2022PRD106.052010}. In the following analysis, given the rather similar masses of $\Sigma_c^+\bar{D}^0$ and $\Sigma_c^+\bar{D}^{*0}$ pairs and the negligible effects of strong decays~\cite{ALICE2022PRD106.052010}, it is assumed that these two systems could be characterized by a common emitting source, which can be first determined by fitting to the $\Sigma_c^+\bar{D}^0$ CF in future experiments and then used to distinguish the spins of $P_c(4440)$ and $P_c(4457)$ by measuring the $\Sigma_c^+\bar{D}^{*0}$ CFs as shown below.

\subsection{Determination of the spins of $P_c(4440)$ and $P_c(4457)$ with the $\Sigma_c\bar{D}^*$ correlation functions}

Although the spins of $P_c(4440)$ and $P_c(4457)$ are not yet known, their masses can be used to determine the $\Sigma_c\bar{D}^{*}$ interaction. In the saturation model adopted here, the interactions of $J^P=(1/2)^-$ and $J^P=(3/2)^-$ are the same, meaning that the cutoff momentum $\Lambda_F$ in the form factor is the only parameter to determine their strengths. We fine-tune $\Lambda_F$ to be $1067$ MeV and $860$ MeV to reproduce the experimental masses of $P_c(4440)$ and $P_c(4457)$, respectively, as shown in Table~\ref{Tab:Pentaquark}. In this work, we refer to the $\Sigma_c\bar{D}^*$ potential as moderately (strongly) attractive if it can dynamically generate the $P_c(4457)$ [$P_c(4440)$] state.

\begin{figure}[htbp]
  \centering
  \includegraphics[width=0.48\textwidth]{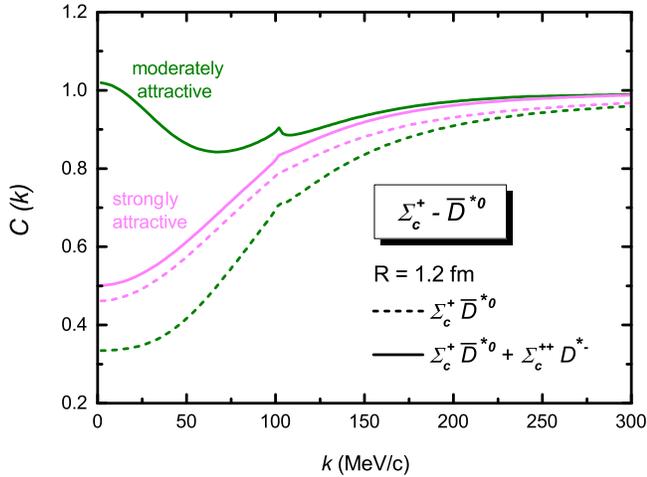}
  \caption{$\Sigma_c^+\bar{D}^{*0}$ correlation function as a function of the relative momentum $k$. The green (light magenta) lines are obtained with the contact potentials capable of generating $P_c(4457)$[$P_c(4440)$]. The source size is set at $1.2$ fm. The dashed line denotes the correlation function which only receives the $\Sigma_c^+\bar{D}^{*0}$ contribution, while the solid line denotes the correlation function to which the $\Sigma_c^+\bar{D}^{*0}-\Sigma_c^{++}D^{*-}$ transition also contributes.
  }\label{Fig:Pc4450_CC}
\end{figure}

Using the so-obtained moderately and strongly attractive $\Sigma_c^+\bar{D}^{*0}$ interactions, we calculate the corresponding CFs. As shown in Fig.~\ref{Fig:Pc4450_CC}, similar to the $\Sigma_c^+\bar{D}^0$ CF, two $\Sigma_c^+\bar{D}^{*0}$ CFs are both suppressed in a wide range of the relative momentum $k$ for $R=1.2$ fm. However, the coupled-channel effect in the CF of the moderate attraction is much larger than that of the strong attraction, which can be traced back to the difference between the binding energies of $P_c(4440)$ and $P_c(4457)$. The opening of the inelastic $\Sigma_c^{++}D^{*-}$ channel leaves a trace in the $\Sigma_c^+\bar{D}^{*0}$ CF around $k\simeq102$ MeV/c. Although the $\Sigma_c^+\bar{D}^{*0}-\Sigma_c^{++}D^{*-}$ coupling strength needs to be further confirmed in a more quantitative way, the qualitative behaviors of the total CFs calculated with the moderate and strong attractions are consistent with the expectations, in which the CF for the shallow bound state is larger than that for the deep bound state for the same source size, as demonstrated in the square-well model~\cite{Liu:2023uly}.

Considering that the spin dependence is not resolved in the standard measurements of CFs, we have to take the spin average for comparison with future experiments. Given that the spins of $P_c(4440)$ and $P_c(4457)$ are not determined, we can define two spin-averaged $\Sigma_c^+\bar{D}^{*0}$ CFs as follows:
\begin{subequations}\label{Eq:averaged}
\begin{align}
  C(k)^A=& \frac{1}{3} C_s + \frac{2}{3} C_m, \\
  C(k)^B=& \frac{2}{3} C_s + \frac{1}{3} C_m,
\end{align}
\end{subequations}
where $C_s$ is the correlation function obtained with the strongly attractive $\Sigma_c\bar{D}^*$ interaction and $C_m$ is that obtained with the moderately attractive interaction, and the superscript $A/B$ denotes scenario A or B. We recall that in scenario A $P_c(4440)$ and $P_c(4457)$ have $J^P=(1/2)^-$ and $J^P=(3/2)^-$, respectively, while scenario B denotes the alternative spin assignments.

\begin{figure*}[htbp]
  \centering
  \includegraphics[width=0.98\textwidth]{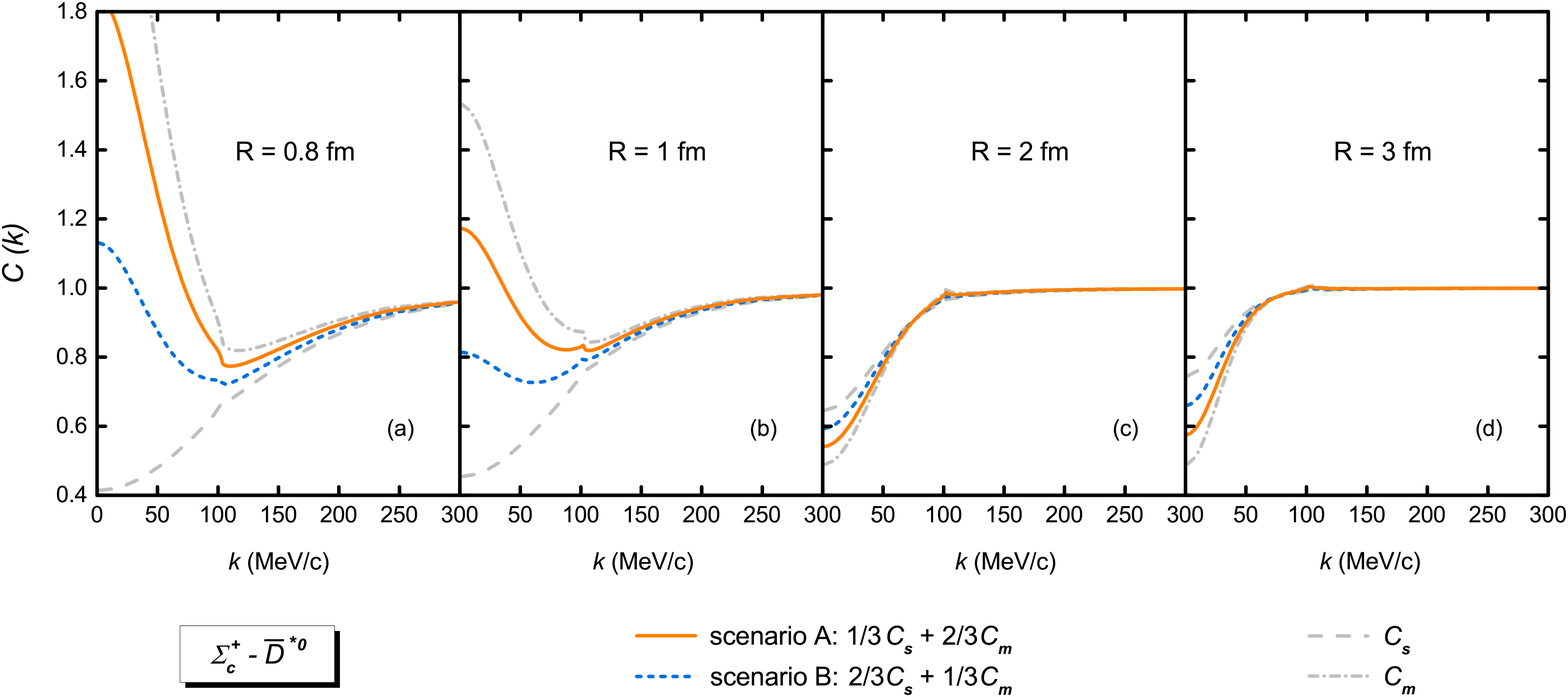}
  \caption{Spin-averaged $\Sigma_c^+\bar{D}^{*0}$ correlation function as a function of the relative momentum $k$ for different source sizes $R=0.8, 1, 2$ and $3$ fm, respectively. The blue short-dashed lines represent the spin-averaged results in scenario A, where $P_c(4440)$ and $P_c(4457)$ have $J^P=(1/2)^-$ and $J^P=(3/2)^-$ respectively, while the orange solid lines represent the results in scenario B where $P_c(4440)$ and $P_c(4457)$ have $J^P=(3/2)^-$ and $J^P=(1/2)^-$. For reference, the $\Sigma_c^+\bar{D}^{*0}$ correlation functions calculated with the moderately (strongly) attractive interactions without specified spin are also shown as the dash-dotted (dashed) lines.
  }\label{Fig:Pc4450_SS}
\end{figure*}

The spin-averaged $\Sigma_c^+\bar{D}^{*0}$ CFs and their source size dependence are shown in Fig.~\ref{Fig:Pc4450_SS}. For reference, we also show the correlation functions calculated with the moderately and strongly attractive interactions without specified spin. We find that the low-momentum behavior of the spin-averaged CF in scenario A is different from that in scenario B. For the small collision systems ($R = 0.8$ and $1$ fm), the results in scenario A are larger than those in scenario B, while the conclusion is reversed for the large collision systems ($R = 2$ and $3$ fm). Especially, the difference in the spin-averaged CFs between scenarios A and B increases with the decreasing source size. The discrepancy can be as large as about $0.7$ in the low-energy limit at $R=0.8$ fm.

Although at first sight, it seems possible to determine the spins of $P_c(4440)$ and $P_c(4457)$ simply from the features of the measured $\Sigma_c^+\bar{D}^{*0}$ CFs, it is actually difficult experimentally. The reason is that  both scenarios A and B can describe the same experimental $\Sigma_c^+\bar{D}^{*0}$ CFs using different source sizes. For example, as shown in Fig.~\ref{Fig:Pc4450_SS} (a) and (b), the spin-averaged $\Sigma_c^+\bar{D}^{*0}$ CF obtained in scenario A with a source size $R = 1.0$ fm is similar to the result obtained in scenario B with a $R = 0.8$ fm. To distinguish these two CFs, high precision is needed experimentally. Thus, it is better to fix the source size from an independent measurement, such as the aforementioned measurement of $\Sigma_c^+\bar{D}^0$ CF in Sec. III. A.

It is worthwhile to note that although all of the above results are obtained with the interactions provided by the resonance saturation model, the conclusion that there exist clear differences between the CFs obtained in the two scenarios is model independent. In fact, the interaction for the partial wave which generates $P_c(4440)$ is always more attractive than the one generating $P_c(4457)$. According to Ref.~\cite{Liu:2023uly}, a moderately attractive interaction and a strongly attractive one may result in completely different low-momentum behaviors in the CFs for smaller emitting sources. This is exactly what we found out here. As shown in Fig~\ref{Fig:Pc4450_SS}, the behavior of the CF of the strong attraction can be classified as one which indicates the existence of a deep bound state while that of the moderate attraction as one which indicates the existence of a shallow bound state. For any other potential, the binding energies of $P_c(4440)$ and $P_c(4457)$ are always input quantities. Therefore, once the source size is determined, as discussed above, one can discriminate the spins of $P_c(4440)$ and $P_c(4457)$ by measuring the $\Sigma_c^+\bar{D}^{*0}$ CFs in $pp$, $pA$, and $AA$ collisions.

\section{Summary}We proposed to discriminate the spins of $P_c(4440)$ and $P_c(4457)$ with femtoscopic correlation functions. We first evaluated the $\Sigma_c\bar{D}^{(*)}$ interactions by reproducing the masses of $P_c(4312)$, $P_c(4440)$, and $P_c(4457)$ in the resonance saturation model, and then calculated the corresponding $\Sigma_c^+\bar{D}^{(*)0}$ correlation functions for the first time. The $\Sigma_c^+\bar{D}^0$ correlation function exhibits an obvious and nonmonotonic source size dependence, which can be used to constrain the size of the emitting source of the $\Sigma_c^+\bar{D}^0$ and $\Sigma_c^+\bar{D}^{*0}$ pairs. Because the low-momentum behaviors of the $\Sigma_c^+\bar{D}^{*0}$ correlation function calculated with a strong $\Sigma_c^+\bar{D}^{*0}$ attraction are significantly different from those calculated with a moderate $\Sigma_c^+\bar{D}^{*0}$ attraction, especially in the case of a small collision system, there are significant differences in the spin-averaged $\Sigma_c^+\bar{D}^{*0}$ correlation functions. Thus, we suggest to determine the spins of $P_c(4440)$ and $P_c(4457)$ by measuring the $\Sigma_c^+\bar{D}^{*0}$ correlation functions in $pp$, $pA$, and $AA$ collisions. It is important to note that the proposed method is model independent, which is determined by the experimental binding energies of $P_c(4440)$ and $P_c(4457)$, as well as the general features of the femtoscopic correlation functions. The method studied in this work can be applied to probe the spin dependence of other hadron-hadron interactions.

\section{Acknowledgments} 
This work is partly supported by the National Natural Science Foundation of China under Grant No. 11975041, and No. 11961141004, and the fundamental Research Funds for the Central Universities. Jun-Xu Lu acknowledges support from the National Natural Science Foundation of China under Grant No. 12105006 and China Postdoctoral Science Foundation under Grant No. 2021M690008. Ming-Zhu Liu acknowledges support from the National Natural Science Foundation of China under Grant No. 12105007 and China Postdoctoral Science Foundation under Grants No. 2022M710317, and No. 2022T150036.

\bibliography{Pc}

\end{document}